\documentclass[12pt]{article}
\usepackage[francais,german,english]{babel}
\usepackage{amssymb}
\def \ub {\underline}
\def \id {1\!\!{\rm I}}

\begin{document}

\thispagestyle{empty}

\selectlanguage{english}

\baselineskip=.5truecm


$\;$

\rightline{{\bf ETH-TH/95-33\quad}}

\rightline{{\bf January '96\quad}}
\vspace{2cm}

\centerline{\bf{\Large{On M-Algebras, the Quantisation of}}}

\smallskip

\centerline{\bf{\Large{Nambu-Mechanics,}}} 

\smallskip

\centerline{\bf{\Large{and Volume Preserving Diffeomorphisms}}}

\vspace{3cm}

{\large
\centerline{{\bf Jens Hoppe \footnote{Heisenberg Fellow. \quad On leave of absence from Karlsruhe University.}}}

\centerline{{\bf Theoretische Physik}}

\centerline{{\bf ETH-H\"onggerberg}}

\centerline{{\bf Schweiz}}}

\vspace{5cm}

\noindent {\bf Abstract:} \ M-branes are related to theories on
 function spaces $\cal{A}$ involving $M$-linear non-commutative maps from
$\cal{A} \ \times \cdots \times \cal{A}$ \ to \ $\cal{A}$. \ While the
Lie-symmetry-algebra of volume preserving diffeomorphisms of $T^M$ cannot be
deformed when $M>2$, the arising $M$-algebras naturally relate to
Nambu's generalisation of Hamiltonian mechanics, e.g. by providing a
representation of the canonical $M$-commutation relations,
$[J_1,\cdots, J_M]=i\hbar$.
Concerning multidimensional integrability, an important generalisation
of Lax-pairs is given. 

\vfill\eject

\section*{1. Introduction}

\noindent Generalizing fundamental concepts, such as Lie algebras or
Hamiltonian 
dynamics, may have quite divers merits; it can lead to new,
interesting possibilities, -- or reassure oneself of our present
notions. While the  result that volume preserving diffeomorphisms of
toroidal $M$-branes, as a Lie-symmetry algebra, cannot be deformed (if
$M>2$) is of the latter nature -- the following ideas appear to be
worthwhile persueing:

\noindent --- Using a $*M$-deformation of the algebra of functions on some
  $M$-dimensional manifold for representing the $M$-linear analogue to
  Heisenberg's commutation relations that Nambu [1]
  encountered in multi-Hamiltonian dynamics.

\noindent --- Generalizing the Jacobi identity  for Lie algebras to a
  (2-bracket) identity involving $2M-1$ elements of a 
  vectorspace $V$ for which an antisymmetric $M$-linear map
  ($M$-commutator) from $V\times \cdots \times V$ to $V$ is defined
  (in a dynamical context, an identity involving $M$,
  rather than 2, of the $M$-commutators, may be preferred).

\noindent --- A potential relevance of $M$-algebras to the quantisation of
  space-time.

\noindent  Perhaps most importantly (on a concrete, practical
  level), an explicit example is given (the
  multidimensional diffeomorphism-invariant integrable field theories
  found in [2]) for the  usefulness (envisaged  some time ago
  [3]) of generalizing Lax-pairs to -triples, \dots .

\vspace{1cm}

\section*{2. M-algebras from M-branes} 

\noindent A relativistic M-brane moving in D-dimensional space time may be
described, in a light-cone gauge,  by the VDiff$\Sigma$-invariant
sector of ([4])
\begin{equation}
H\;=\;\frac 1 2 \;\int_\Sigma\;\frac{d^M\varphi}{\rho (\varphi)}\;(\vec
p^{\;2} + g)
\end{equation}
where $g$ is the determinant of the M$\times$M matrix $(g_{rs}) :=
(\nabla_r x^i \nabla_s x_i)_{r,s=1\cdots M}$, \ $x^i$ and $p_i$\break
$(i=1,\cdots, D-2 =: d)$ are canonically conjugate fields, and $\rho$
is a fixed non-dynamical density on the $M$-dimensional
parameter-manifold $\Sigma$ \ ($M=1$ for strings, $M=2$\ for
membranes,\dots). \ 
Generators of VDiff$\Sigma$, the group of volume-preserving
diffeomorphisms of $\Sigma$ (resp. the component connected to the
identity), are represented by
\begin{equation}
K\; :=\; \int_\Sigma \; f^r\; p_i\; \partial_r\; x^i \; d^M\;\varphi
\end{equation}
with $\nabla_r f^r=0$. \ $g$ may be written as
\begin{equation}
g\;=\;\sum_{i_1<i_2<\cdots<i_M}\{ x_{i_1} ,\cdots, x_{i_M}\}
\{x^{i_1}, \cdots, x^{i_M}\},
\end{equation}
where the `Nambu-bracket' $\{\cdots\}$ is defined for functions
$f_1,\cdots,f_M$ on $\Sigma$ as 
\begin{equation}
\{f_1,\cdots,f_M\}\; :=\; \epsilon^{r_1\cdots r_M}\;\partial_{r_1}\;f_1
\cdots \partial_{r_M}\;f_M.
\end{equation}
This trivial, but important observation suggests to consider
Hamiltonians 
\begin{equation}
H_\lambda \;:=\;\frac 1 2 \; Tr \biggl( \vec P^{\;2} \pm
\sum_{i_1<\cdots<i_M} [X_{i_1},\cdots,X_{i_M}]_\lambda^2\biggr),
\end{equation}
resp.
\begin{equation}
H_\lambda\;=\;\frac 1 2\;\sum_{i=1}^d \beta \;(P_i,P_i)\;+\;\frac 1
2\; \sum_{i_1<\cdots<i_M} \beta \;\bigl(
[X_{i_1},\cdots,X_{i_M}]_\lambda, [X_{i_1},\cdots X_{i_M}]_\lambda\bigr),
\end{equation}
where $X^i$ and $P_i$ are elements of (possibly finite dimensional,
$\lambda$-dependent) vectorspaces $V$ on which antisymmetric $M$-linear
maps $[,\cdots,]_\lambda\;:\;V \times\cdots \times V \to V$ are defined, and
$\beta$ a positive definite hermitean form, preferably invariant with
respect to some analogue of volume preserving diffeomorphisms
(cp. (2)). 

With
\begin{equation}
[T_{a_1},\cdots,T_{a_M}]_\lambda\;=\;f_{a_1\, \cdots\, a_M}^a\;(\lambda)\;T_a
\end{equation}
and
\begin{equation}
\beta (T_a, T_b)\;=\;\delta_b^a
\end{equation}
for some (possibly $\lambda$-dependent) basis \ $\{ T_a\}_{a=1}^{{\rm
    dim} V}$ \ of $V$, i.e.
\begin{equation}
f_{a_1\cdots a_M}^a (\lambda)\;=\;\beta\bigl( T_a,
[T_{a_1},\cdots,T_{a_M}]_\lambda\bigr)\,,
\end{equation}
(6) reads
\begin{eqnarray}
H_\lambda\;=\;\frac 1 2 \; p_{ia}^* p_{ia}&+&\frac 1 2 \; \bigl(
f_{a_1\cdots a_M}^a (\lambda)\bigr)^*\; f_{b_1\cdots b_M}^a
(\lambda)\nonumber\\
&& \frac{1}{M!}\;x_{i_1 a_1}^* \cdots x_{i_M a_M}^*\;x_{i_1b_1} \cdots
x_{i_Mb_M}\;  ,
\end{eqnarray}
while (1) may be written as
\begin{eqnarray}
H\;=\;\frac 1 2\; p_{i\alpha}^* p_{i\alpha}&+&\frac 1 2 \;\bigl(
g_{\alpha_1\cdots \alpha_M}^\alpha\bigr)^*\;
g_{\beta_1\cdots\beta_M}^\alpha\nonumber\\
&& \frac{1}{M!}\; x_{i_1\alpha_1}^* \cdots x_{i_M \beta_M}\ ;
\end{eqnarray}

\begin{equation}
g_{\alpha_1\cdots\alpha_M}^\alpha\; :=\; \int_\Sigma Y_\alpha^* \bigl\{
Y_{\alpha_1},\cdots,Y_{\alpha_M}\bigr\}\; \rho\;d^M\;\varphi
\end{equation}
is defined with respect to some orthonormal basis of functions (on
$\Sigma$) satisfying
\begin{eqnarray}
\int Y_\alpha^*\;Y_\beta\;\rho\;d^M\varphi&=&
\delta_\beta^\alpha\nonumber\\
&& \alpha,\beta\;=\;1 \cdots \infty
\end{eqnarray}
(even for real $x_i$, it is often convenient to take a complex basis).

\noindent Obvious questions are:
\begin{description}
\item[1)] Does there exist a `natural' sequence of finite dimensional
  vectorspaces $V_n$ with basis $\{ T_a^{(n)}\}$ \ and 
  antisymmetric maps \ $F_n : V_n \; \times\cdots \times \; V_n \to V_n$ \ such
  that for each $(M+1)$-tuple \ $(a\;a_1\cdots a_M)$
\begin{equation}
\lim_{n\to\infty}\;f_{a_1\cdots a_M}^a\;(\lambda_n)\;
\displaystyle\mathop{?}_=\; g_{a_1\cdots a_M}^a\; .
\end{equation}
\item[2)] For which $M$ do there exist finite dimensional analogues of
  (2), $K(n)$, leaving $(10)_{\lambda_n}$ \ invariant,
  such that, as $n\to\infty$, the full invariance under
  volume-preserving diffeomorphisms is recovered?
\item[3)] What about $\lambda$-deformations with infinite dimensional
  $V$'s ?
\end{description}
Let us look at the case of a $M$-torus, $\Sigma=T^M$ :

\noindent Choosing 
\begin{equation}
Y_{\vec m}\;=\;e^{i\,\vec m \,\vec
  \varphi}, \; \vec m \;=\; (m_1, \cdots,
m_M)\;\in\;\mathbb{Z}^M,\;\rho\;\equiv\;1,
\end{equation}
one gets
\begin{equation}
g_{\vec m_1 \cdots \vec m_M}^{\vec m}\;=\;i^M (\vec m_1, \cdots, \vec
m_M)\; \delta_{\vec m_1 +\cdots +\vec m_M}^{\vec m}
\end{equation}
where \ $(\vec m_1,\cdots, \vec m_M)\; \in \;\mathbb{Z}$ \ denotes the
determinant of the corresponding \ $M\times M$ Matrix (an element of
$GL(M,\mathbb{Z}$) ).

Consider now the following `$*M$-product' (a deformation of the
ordinary commutative product of $M$ functions $f_1,\cdots,f_M$ on
$\Sigma$):
\begin{eqnarray}
(f_1\cdots f_M)_*\;:=\; f_1\cdots f_M 
&+&\sum_{m=1}^\infty
\begin{array}{cc}
\bigl(\frac{(-i)^{M+1}\lambda}{M!}\bigr)^m &\epsilon^{r_1r_1'\cdots
  r_1^{(M)}}\\
\frac{\qquad}{m!}\quad \cdots &\epsilon^{r_mr_m'\cdots
  r_m^{(M)}}
\end{array}\nonumber\\
&& \qquad \frac{\partial^m \,f_i}{\partial\varphi^{r_1}\cdots
  \partial\varphi_{r_m}}\; \cdots\;
\frac{\partial^m\,f_M}{\partial\varphi^{r_1^{(M)}}\cdots
  \partial\varphi^{r_m^{(M)}}}\ .
\end{eqnarray}
One then finds that
\begin{eqnarray}
\bigl( Y_{\vec m_1} \cdots Y_{\vec m_M}\bigr)_* &=&
\sqrt{\omega}^{\,-\,(\vec m_1,\cdots,\vec m_M)}\; Y_{\vec m_1 + \cdots
  \vec m_M}\nonumber \\
&& \sqrt{\omega}\;=\; e^{i\,\frac{\lambda}{M!}}\; .
\end{eqnarray}
Defining
\begin{equation}
[f_1,\cdots,f_M]_*\;:=\;\sum_{\sigma \;\in\;
  S_M}\;({\rm sign}\;\sigma)\;(f_{\sigma\,1}\cdots f_{\sigma\,M})_*
\end{equation}
to simply be the antisymmetrized $*M$-product, one gets
\begin{equation}
[T_{\vec m_1},\cdots, T_{\vec m_M}]\;=\;\frac{-i}{2\pi\Lambda}\;\sin\;
\bigl(2\pi\Lambda\,(\vec m_1,\cdots,\vec m_M)\big)\; T_{\vec
  m_1\,+\cdots+\,\vec m_M}
\end{equation}
\[
{\rm with \ }\Lambda \;:=\;\frac{\lambda}{2\pi M!}\quad{\rm and}\quad
T_{\vec m}\;:=\;\lambda^{-\;\frac{1}{M-1}}\;Y_{\vec m}.
\]


For $M>1$ arbitrary (but fixed), let $V$ denote the vectorspace (over
$\mathbb{C}$) generated by \ $\{T_{\vec m}\}_{\vec m \;\in\;
  \mathbb{Z}^M}$,\ \ $\mathbb{M}^\Lambda$ denote $(V,*)$ and
$\mathbb{A}^\Lambda$ \ denote $(V,[\cdots]_*)$. 

The hermitean form $\beta$ (cp. (8),(9)),
\[
\beta\,(T_{\vec m},T_{\vec n})\;=\;\delta_{\vec n}^{\vec m},\enskip 
\;\beta (c_i\,X_i,d_j\,X_j)\;=\;c_i^*\,d_j\;\beta (X_i, X_j),
\]
will have the important property (`invariance') that (for \
$X_i=x_{i\vec m} T_{\vec m}$ \ with \ $x_{i\vec m}^*=x_{i-\vec m}$)
\[
\beta\, \left(X, [X_{i_1}, \cdots X_{i_M}]\right)\;=\;-\,\beta \left(
  X_{i_r},[X_{i_1},\cdots, X_{i_{r-1}}, X, X_{i_{r+1}},\cdots,
  X_{i_M}]\right), 
\]
as
\[
\beta\,\left( T_{\vec m}, [ T_{\vec m_1},\cdots, T_{\vec m_M}]\right)
\;=\; \frac{-\;i}{2\pi\Lambda}\;\delta_{\vec m_1, +\cdots +\vec
  m_M}^{\vec m} \; \sin \left( 2\pi\Lambda (\vec m_1, \cdots, \vec
  m_M)\right). 
\]


For rational \ $\Lambda = \frac{\widetilde N}{N}$ \ (assuming $N$ and
$\widetilde N<N$ \ having no common divisor $>1$) both \
$\mathbb{A}^\Lambda$ \ and \ $\mathbb{M}^\Lambda$ \ may be divided by
an ideal of finite codimension, namely (using the periodicity of the
structure-constants) the vectorspace I generated by all elements of
the form \ $T_{\vec m} - T_{\vec m + N\ {\rm (anything)}}$.  One thus arrives
at considering (for arbitrary odd $N$)
\begin{equation}
V^{(N)}\;:=\;\bigg\langle T_{\vec
  m}|m_r\;=\;-\;\frac{N-1}{2}\;,\cdots,\;+\;\frac{N-1}{2}
\bigg\rangle_{\mathbb{C}}\; \quad r\;=\;1\cdots M  
\end{equation}
with a \ $*_M$ product on $V^{(N)}$ defined just as in (18):
\begin{eqnarray}
&&(T_{\vec m_1}\cdots T_{\vec m_M})_*\;:=\;\frac{-\,i\,N}{2\pi\widetilde N
  M!}\;\omega^{-\,\frac 1 2\;(\vec m_1,\cdots,\vec m_M)} \;
T_{\vec m_1  +\cdots+\vec m_M\; ({\rm mod\ } N)}  \nonumber\\
&&\omega\;=\;e^{4\pi\,i\,\frac{\widetilde N}{N}}\; ,
\end{eqnarray}
and a corresponding alternating product,
\begin{eqnarray}
&&[T_{\vec m_1},\cdots, T_{\vec m_M}]_*\;=\;\frac{-\,i\,N}{2\pi\widetilde
    N}\;{\rm sin}\;\bigl( 2\pi\;\frac{\widetilde N}{N}\;(\vec m_1, \cdots, \vec
  m_M)\bigr)\; T_{\vec m_1 +\cdots+\vec m_M}\;_{({\rm mod\ } N)}  \nonumber \\
&&\vec m_r\;\in\;(\mathbb{Z}_N)^M \; . 
\end{eqnarray}
The `structure constants' of the alternating finite dimensional
$M$-algebras
\begin{eqnarray}
&&\mathbb{A}_N\;:=\;\bigl( V^{(N)}, [,\cdots,]_*\bigr),\nonumber\\
&&f_{\vec m_1\cdots\vec m_M}^{(N)\,\vec
  m}\;:=\;\frac{-\,i\,N}{2\pi\widetilde N}\;{\rm
  sin}\;\biggl(2\pi\;\frac{\widetilde N}{N}\;\biggl(\vec m_1,\cdots,\vec
m_M\biggr)\biggr)\;\cdot\;\delta_{\vec m_1 +\cdots+ \vec m_M}^{\vec m}\;_{({\rm mod\ } N)}
\end{eqnarray}
satisfy (14) (up to an $N$ and $\mathbb{Z}_N^M$-independent rescaling
of the generators, resp. factors of $i$, which anyway drop out in
(10) and (11); \ $n=N^M$, $f^{(N)}\;\stackrel{\textstyle\wedge}{=}\;
f(\lambda_n)$, $\vec m\;\in\;\mathbb{Z}_N^M$ $V^{(N)}\;=\;V_{n=N^3}$, and
$\displaystyle\mathop{\lim}_{N\to\infty} \;V^{(N)} = V$).
\begin{eqnarray}
H_N&=& \frac 1 2 \; p_{i-\vec m}\;p_{i\vec m} \nonumber\\
&+& \frac 1 2\; \frac{N^2}{4\pi^2\;\widetilde N^2}\;{\rm sin}\;\biggl(
2\pi \frac{\widetilde N}{N}\;\biggl(\vec m_1 \cdots \vec
m_M\biggr)\biggr)\;\cdot\; 
{\rm sin}\;\biggl( 2\pi\;\frac{\widetilde N}{N}\;\biggl(\vec n_1, \cdots \vec
n_M\biggr)\biggr)\nonumber\\
&&\frac{1}{M!}\;\cdot\;x_{i_1-\vec m_1} \cdots x_{i_M-\vec m_M} \;
x_{i_1\vec n_1} \cdots x_{i_M \vec n_M}\; \delta_{\vec n_1 +\cdots +
  \vec n_M}^{\vec m_1 +\cdots +\vec m_M}\;_{({\rm mod\ } N)}
\end{eqnarray}
could therefore be considered as a finite-dimensional analogue of (1).

\vspace{1cm}

\section*{3. Multidimensional Commutation Relations}

\noindent Before turning to questions of symmetry, let me discuss in a
little more detail the \break 
$*M$-algebras $\mathbb{M}^\Lambda$, with
defining relations (cp.~(18); note the slight change of
notat\-ion/normalisation)
\[
\bigl(T_{\vec m_1}\cdots T_{\vec m_M}\bigr)_*\;=\;\omega^{-\,\frac 1
  2\;(\vec m_1, \cdots, \vec m_M)}\; T_{\vec m_1 +\cdots +\vec m_M}\;(*)\;,
\]
and as vectorspaces generated by basis-elements \ $T_{\vec m}, \ \vec
m\in S$ \ (where $S=\mathbb{Z}^M$, \ $S=(\mathbb{Z}_N)^M$, \
 \ or any combination thereof -- in the $M$-brane
context, depending on whether $\Sigma=T^M$, resp. a fully, or
partially,  discretized $M$-torus).

First of all note, that for any $M$ elements, \ $A_1,\cdots A_M \in
V$, any even permutation $\sigma \in S_M$ (the symmetric group in $M$
objects), and any choice of $S$ (even $\mathbb{R}^M$),
\begin{equation}
(A_1\cdots A_M)_*\;=\;(A_{\sigma(1)}\cdots A_{\sigma(M)})\quad ({\rm
  sign}\;\sigma \;=\;+)\;,
\end{equation}
and that $E := T_{\vec 0}$ acts as a `unity' in the sense that if one
of the $A_r$ is equal to $T_{\vec 0}$, the $*M$-product becomes
commutative (i.e. independent of the order of its $M$ entries).

Using $E$, one may identify $T_{(m=\pm |m|,0,\cdots, 0)}$ with the $|m|$-th
power of $E_{\pm 1} := T_{(\pm 1,0,\cdots,0)}$,
\begin{eqnarray}
T_{(m,0,\cdots,0)}&=&\left(\bigl( (( E\cdots EE_{\pm 1})_*\cdots
EE_{\pm 1})_*\cdots\bigr)_*\cdots EE_{\pm 1} \right)_*\;,\\
&&\quad \uparrow\nonumber\\
&& |m|\;{\rm brackets}\nonumber
\end{eqnarray}
so that one may wonder whether $\mathbb{M}^\Lambda$ can be thought of
as being generated by
\[
E\;=\;T_{\vec 0},\;E_{\pm 1}\;=\;
T_{(\pm 1 \ 0\cdots0)}, \cdots, \;
E_{\pm M}\;=\;
T_{(0\cdots 0\;\pm 1)}\; .
\]
This is indeed the case: \  Let $\mathbb{F}^M$ be the free (non
associative) $M$-algebra generated by $2M+1$ elements $E,
E_{\pm 1},\cdots,E_{\pm M}$; define arbitrary powers $(E_r)^m$ of the generating
elements as in (27) (from now on \ $E_{-r}^{|m|} =: E_r^{- |m|}$, \ a
notation that will be justified via (29)), and let
\begin{equation}
E_{\vec m}\;:=\;E_1^{m_1}\;E_2^{m_2}\;\cdots\;E_M^{m_M}\; .
\end{equation}
Divide $\mathbb{F}^M$ by the ideal generated by elements 
\begin{equation}
E_{\vec m'}\;E_{\vec m''} \cdots E_{\vec m^{(M)}}\;-\;
\omega^{\gamma(\vec m', \vec m'',\cdots, \vec m^{(M)})}\;\cdot\;
E_{\vec m' + \cdots + \vec m^{(M)}}
\end{equation}
where $\omega = e^{4\pi\,i\,\Lambda}$ and
\begin{eqnarray}
2\gamma (\vec m',\cdots,\vec m^{(M)})&:=& (m_1\cdot m_2\cdot\; \cdots\;
\cdot m_M)\;-\; (\vec m', \vec m'', \cdots, \vec m^{(M)})\nonumber \\
&-& \sum_{r=1}^M \; \biggl( \prod_{s=1}^M \; m_s^{(r)}\biggr) \\
 (\vec m\;:=\;\vec m'\;+\;\vec m''&+&\cdots \; +\;\vec m^{(M)})\;. \nonumber
\end{eqnarray}
This quotient then is isomorphic to $\mathbb{M}^\Lambda$, as can be
seen by defining
\begin{equation}
T_{\vec m}\;:=\;\omega^{\frac 1 2\;m_1m_2\cdots m_M}\;
E_1^{m_1}\;E_2^{m_2}\cdots E_M^{m_M}\; ,
\end{equation}
which (due to (29) being zero in $\mathbb{F}^\Lambda / I$) satisfies
(18) (with $Y$ standing for $T$). 

Note that 
\begin{equation}
E_2^{m_2}\;E_1^{m_1}\;E_3^{m_3}\;\cdots\;E_M^{m_M}\;=\;\omega^{m_1m_2\cdots
  m_M}\;\cdot\;E_1^{m_1}\;E_2^{m_2}\cdots 
E_M^{m_M},
\end{equation}
in particular:
\begin{equation}
E_2\,E_1\,E_3\;\cdots\;E_M\;=\;\omega\;E_1\,E_2\;\cdots\;E_M
\end{equation}
(while any even permutation does not alter the product, cp. (26)).

In order to get a feeling for (29)/(30) it may be instructive to
consider $M=3$: due to (29),
\begin{eqnarray}
&&(E_1^{n_1}\;E_2^{n_2}\;E_3^{n_3})(E_1^{l_1}\;E_2^{l_2}\;E_3^{l_3})
(E_1^{k_1}\;E_2^{k_2}\;E_3^{k_3})\nonumber\\ 
=&& E_1^{n_1+l_1+k_1}\;E_2^{n_2+l_2+k_2}\;E_3^{n_3+l_3+k_3}\nonumber\\
&&\cdot \; \omega^{n_1l_3k_2\,+\,n_2l_1k_3\,+\,n_3l_2k_1}\\
&&\cdot \;
\sqrt{\omega}^{\;n_1(l_2l_3\,+\,k_2k_3)\,+\,n_2(l_1l_3\,+\,k_1k_3)\,+\,
  n_3(l_1l_2\,+\,k_1k_2)}\nonumber\\
&&
\cdot\;\sqrt{\omega}^{\;n_1n_2(l_3\,+\,k_3)\,+\,n_1n_3(l_2\,+\,k_2)\,+\,n_2n_3
  (l_1\,+\,k_1)}\quad .\nonumber
\end{eqnarray}

The general rule (30) can hence be stated as follows:

Consider all possible triples (resp. $M$-tuples) containing powers of
each of the $E_r(r=1\cdots M)$ exactly once. If the `contraction'
picks out exactly one factor from each of the 3 (resp. $M$) factors in
(34) it does \underbar{not} contribute if they are already in the
correct order, modulo even permutations (cp.~26), (like \
$E_1^{n_1}\,E_2^{l_2}\,E_3^{k_3}$, or
$E_2^{n_2}\,E_3^{l_3}\,E_2^{k_1}$), while they contribute
a factor $\omega^{{\rm (product\  of\ the}\;E{\rm -powers)}}$, when they
are \ub{not} 
in the correct (modulo even permutation) order (like
$E_2^{n_2}\,E_1^{l_1}\,E_3^{k_3}$). Contractions entirely within one
of the factors don't contribute, while mixed contractions (involving
at least 2, but not all, of the factors in (34)), all contribute a
factor $\sqrt{\omega}^{{\rm (product \ of\ the}\;E{\rm -powers)}}$,
irrespective of their order. 

Due to (32), all `monomials' are proportional to one of the elements
$E_{\vec m}$ (cp.~(28)) -- which therefore form a basis (with the
convention $E_{\vec 0} \equiv E$). Note that
$2\pi M!\,\Lambda=\lambda\to 0$ is a `classical limit' (resp. $\lambda
\neq 0$ a `quantisation' of the classical Nambu-structure) as,
formally, 
\begin{equation}
[\ln\,E_1, \ln \,E_2,\,\cdots ,\,\ln\,E_M]\;=\;i\,\lambda\;E\; .
\end{equation}
Having obtained this relation, one could of course start with objects \
$\ln E_r=: J_r$, $[J_1,J_2$,$\cdots,J_M]=i\,\lambda\,E$, and derive
generalized `Hausdorff-formulae' for products involving the
$e^{i\,m_r\,J_r}$.

Of course, (35) cannot be true in any $M$-algebra containing only
finite linear combinations of the basis-elements $E_{\vec m}$, as
$T_{\vec 0}=E$ never appears on the r.h.s. of (20); this is similar to
the fact that the canonical commutation relations of ordinary quantum
mechanics, $[q,p] =i\, \hbar\;\id$, cannot hold for trace-class
operators. (35) may be justified by formally expanding
$\ln\,E_r\;=\;-\displaystyle\mathop{\Sigma}_{n_r=1}^\infty\;
\displaystyle\mathop{\Sigma}_{k_r=0}^{n_r}\;{n_r \choose
  k_r}\;\frac{(-)^{k_r}}{n_r}\;E_r^k$, \ using
\[
[E_1^{k_1},E_2^{k_2},\cdots,E_M^{k_M}]\;=\;\frac{M!}{2}\;(1-\omega^{k_1\cdots
  k_M})\;E_1^{k_1}\cdots E_M^{k_M}
\]
and then resumming recursively, after the first step obtaining
\begin{equation}
\frac{M!}{2}\ln E_1\cdots \ln E_M\,-\,\frac{M!}{2}\sum_{n_r,k_r
  \atop r>1}\,'\cdots \ln (E_1 \omega^{k_2\cdots
  k_M})E_2^{k_2}\cdots E_M^{k_M}\,=\,\frac{M!}{2} (\ln
\omega)\,\cdot\,E\; ,
\end{equation}
as formally,
\[
\ \sum_{n_r=1}^\infty\;\sum_{k_r=1}^{n_r}\;{n_r \choose k_r}\;
\frac{(-)^{k_r}}{n_r}\;k_r\;E_r^k\;=\;E_r\;\cdot\;
\sum_{n'=0}^\infty\;(E-E_r)^{n'}\;=\;E\;. 
\]

\vspace{1cm}

\section*{4. Breakdown of Conventional Symmetries}

\noindent Let us now discuss the question, whether theories like (5)
or (6) can have symmetries reminiscent of volume preserving
diffeomorphisms; in particular whether the generators (2) may be
`translated' to finite dimensional
analogues.
\renewcommand{\thefootnote}{\fnsymbol{footnote}}\footnote[1]{{\normalsize  For $M=2$, this question
was already considered in [4] and answered positively.}} \renewcommand{\thefootnote}{\arabic{footnote}} For simplicity, consider
again $\Sigma = T^M$. 

As \ $f^r = \partial_s\omega^{rs}=\epsilon^{rsr_1\cdots r_{M-2}}
\partial_s \omega_{r_1\cdots r_{M-2}}$ \ for non-constant
(divergence-free) vector-fields on $T^M$, (2) may be written in the
form
\begin{equation}
K_{r_1\cdots r_{M-2}}\;=\;\int d^M\varphi\;\omega_{r_1\cdots
  r_{M-2}}\,\bigl\{ p_i, x^i, \varphi^{r_1},\cdots,\varphi^{r_{M-2}}\bigr\}\;,
\end{equation}
resp., in Fourier-components,
\begin{equation}
K_{r_1\cdots r_{M-2}}^{\vec l}\;=\;\sum_{\vec m,\vec n \atop
\in\,\mathbb{Z}^M} \; \delta_{\vec m +\vec n}^{\vec l}\;p_{i\vec
m}\,x_{i\vec n}\,(\vec m, \vec n, \vec e_{r_1},\cdots,\vec e_{r_{M-2}})
\end{equation}
(where $\vec e_r$ denotes the unit vector in the $r$-direction). 

Suppose the deformed theory was invariant under transformations that
are still generated in a conventional way by phase-space functions of
the form
\begin{equation}
K^{\vec l}\;=\;\sum_{\vec m, \vec n \,\in\,S}\;p_{i\vec m}x_{i\vec
  n}\;\delta_{\vec m +\vec n}^{\vec l}\;c_{\vec m\vec n}\; .
\end{equation}
Using $[x_{i\vec m}, p_{j\vec n}]=\delta_{ij} \delta_{\vec m}^{-\vec
  n}$, while leaving open whether $S=\mathbb{Z}^M$ or
$S=(\mathbb{Z}_N)^M$ as well as (independently) whether $\delta$ is
defined \ mod $N$, or not, one has
\begin{eqnarray}
&&[K^{\vec l}, \widetilde{K}^{\vec l\,'}]\;=\;\sum_{\vec m_1 \vec n
  \atop \in\,S}\;p_{i\vec m} x_{i\vec n}\;\delta_{\vec m + \vec
  n}^{\vec l + \vec l\,'}\;{\stackrel{\textstyle\approx}{c}}_{\vec m
    \vec n}\nonumber\\
&& {\rm with} \\
&& {\stackrel{\textstyle\approx}{c}}_{\vec m
    \vec n} \;=\;\sum_{\vec k\,\in\,S}\;\Biggl(\delta_{\vec k}^{\vec l
    - \vec m}\; \delta_{- \vec k}^{\vec l\,'-\vec n}\;
c_{\vec m \vec k}\;
  \tilde c_{-\vec k \vec n}\,-\,{\vec l \leftrightarrow \vec l\,'\choose c
    \leftrightarrow \tilde c} \Biggr)\; ,\nonumber
\end{eqnarray}
while $\dot{K}^{\vec l} = 0$ would require $c_{\vec m \vec n} =
- -\,c_{\vec n \vec m}$ \ and
\begin{eqnarray}
&&\sin \left(2\pi\Lambda (\vec a_1,\cdots, \vec a_M)\right)\;
\sin\left( 2\pi\Lambda(\vec a_1+\cdots +\vec a_M, \vec a_2\,',\cdots ,\vec
a_M\,')\right)\nonumber\\
&&\cdot\; c_{\vec a_1 + \cdots \vec a_1\,' + \cdots \vec a_M\,', \vec
  a_1\,'}\;\cdot\; x_{i_1\vec a_1}\;x_{i_1\vec a_1\,'} \; \cdots \; x_{i_M\,
\vec a_M}\;x_{i_M\,\vec a_M\,'}\;=\;0
\end{eqnarray}
(where for (41) consistency of the $\delta$-functions used in (39) and
(25)$_\Lambda$ with the index set $S$ was assumed).

The effect of the $x_{i\vec m}$-factors in (41) is to make the product
$\sin \,\cdot\,\sin\,\cdot\,c$, symmetric under any interchange \
$\vec a_r \leftrightarrow \vec a_r\,'$, as well as any simultaneous
interchange \ $\vec a_r \leftrightarrow \vec a_s$, $\vec a_r\,'
\leftrightarrow \vec a_s\,'$. Choosing, e.g., $\vec a_r\,' = \vec a_r
(r=1\cdots M)$, with \ $\sin (2\pi\Lambda (\vec a_1\cdots \vec
a_M))\neq 0$, (41) requires that
\begin{equation}
\sum_{\sigma\,\in\,S_M}\;c_{\vec a_{\sigma 1} \,+\,2 (\vec a_{\sigma 2}
    + \cdots + \vec a_{\sigma M}), \vec a_{\sigma 1}}\;=\;0\;.
\end{equation}
This condition is insensitive to any alteration of the deformation:
replacing the sine-function in (41) (resp.~(25)$_\Lambda ,\cdots$) by
any other function of the determinant will still result in (42) as a
necessary condition. Apart from $M=2$ $(c_{\vec a_1+2\vec a_2,\vec
  a_1}+\;c_{\vec a_2 + 2\vec a_1, \vec a_2}= 0$ \ is trivially
satisfied by any odd function) (42) is \ub{not} satisfied by
\begin{equation}
c_{\vec m \vec n}\;=\;\sin \bigl( 2\pi\Lambda (\vec m, \vec n, \vec
k_1, \cdots, \vec k_{M-2})\bigr)\;,
\end{equation}
- -- nor would (40) be a linear combination of the generators (39), for
such a $c_{\vec m \vec n}$; \ for $M=3$, e.g., one would obtain
\begin{eqnarray}
&&{\stackrel{\textstyle\approx}{c}}_{\vec m
    \vec n} (\vec l \,\vec l\,'; \vec k \,\vec k\,')\nonumber\\
&=&\sin \biggl( 2\pi\Lambda\,\biggl(\vec l, \,\vec l\,',\frac{\vec k + \vec
  k\,'}{2}\biggr)\biggr) \nonumber\\
&&\quad \cdot\;\sin\,\biggl(2\pi\Lambda\,\biggl(\biggl(\vec m, \vec
n,\;\frac{\vec k + \vec k\,'}{2}\biggr)\,+\,\biggl(\vec m - \vec n,
\frac{\vec l - 
  \vec l\,'}{2},\;\frac{\vec k - \vec k\,'}{2}\biggr)\biggr)\biggr)\\
&-& \sin \biggl(2\pi\Lambda\,\biggl(\vec l,\vec l\,',\;\frac{\vec k-\vec
  k\,'}{2}\biggr)\nonumber\\
&&\quad \cdot \sin \biggl( 2\pi\Lambda\,\biggl(\vec m,\vec
n,\;\frac{\vec k - \vec k\,'}{2}\biggr)\,+\,\biggl(\vec m - \vec
n,\;\frac{\vec l - \vec l\,'}{2},\;\frac{\vec k+\vec
  k\,'}{2}\biggr)\biggr)\nonumber 
\end{eqnarray}
- -- which means that the algebra closes only for \ $\vec k\,'=\vec k$
(for $\Lambda = \frac 1 N$ \ this would give $N^3$ closed Lie
algebras, each of dimension $N^3$; \ in fact, each consisting of $N$
copies of \ $gl(N)$). -- In any case, if $c_{\vec m \vec n}$ was a
function of \ $(\vec m_1 \vec n_1 \vec k_1, \cdots, \vec k_{M-2})$,
one could let $\vec a_2, \vec a_3, \cdots \vec a_M$ differ only in the
(`irrelevant') $\vec k_1, \cdots \vec k_{M-2}$ directions and obtain
\begin{equation}
f\;\bigl(\bigl((2M-2)\vec a_2, \vec a_1,\cdots \bigr)\bigr)
\;+\;(M-1)\,f\,\bigl((2\vec a_1, \vec a_2, \cdots ) \bigr)\;=\;0\;,
\end{equation}
which eliminates all \ $c_{\vec m \vec n}$ that are non-linear
functions of the determinant.


Interestingly, $c_{\vec m \vec n}=(\vec m, \vec n,$ something)$_{{\rm
  if}\ M>2}$ \ is suggested by yet another consideration: \ replacing
\ $\{ p_i, x_i, \varphi^3, \cdots, \varphi^M\}$ \ (cp.~(37); for
notational simplicity taking $r_1=3,\cdots, r_{M-2}=M$) \ by
\begin{equation}
[P_i, X_i, \ln E_3, \cdots, \ln E_M]\; ,
\end{equation}
(with \ $P_i = p_{i\vec m} T_{\vec m}, X_i=x_{i\vec m} T_{\vec m})$
formally expanding the logarithms in a power series, using (20), and
then (formally) summing again, one obtains something proportional to 
\begin{equation}
p_{i\vec m}\,x_{i\vec n}\; T_{\vec m +\vec n}\;\cdot\;(m_1\,n_2 -
m_2\,n_1)\;.
\end{equation}
\begin{eqnarray}
&& [P_i, X_i, \ln E_3, \cdots, \ln E_M]\nonumber\\
&=& p_{i\vec m} \;x_{i\vec n}\;(-)^{M-2} \sum_{n_3=1}^\infty
\sum_{k_3=0}^{n_3}\cdots \sum_{n_M=1}^\infty \sum_{k_M=0}^{n_M} { n_3
  \choose k_3}\cdots {n_M \choose k_M}\; \frac{(-)^{k_3+\cdots
    +k_M}}{n_3\cdots n_M}\nonumber\\
&&\qquad \cdot\; [ T_{\vec m}, \;T_{\vec n},\; E_3^{k_3}, \cdots,
E_M^{k_M}]\nonumber\\
&\sim&\sum\!\!\!\cdots \sin\left(2\pi\Lambda\,(\vec m, \vec n, k_3\, \vec
  e_3,\cdots, k_M\,\vec e_M)\right)\;\cdot\; T_{\vec m + \vec n + \vec
  k}\nonumber \\
&\sim& \sum\!\!\!\cdots\left(\sqrt{\omega}^{\;k_3\cdots k_M\;z} -
  \sqrt{\omega}^{\;- k_3\cdots
    k_M\;z}\right)\;(\sqrt{\omega})^{\prod_{r=1}^M
  (m_r+n_r+k_r)}\;\cdot \nonumber\\
&&\qquad \cdot \; E_1^{m_1+n_1}\; E_2^{m_2+n_2}\;
E_3^{m_3+n_3+k_3}\;\cdots\; E_M^{m_M+n_M+k_M}\nonumber \\
&\sim& \sum^{\qquad'}\!\!\!\!\!\cdots\biggl( \ln \biggl( \sqrt{\omega}^{\;k_4\cdots
    k_M\;z\,+\,\prod_{r\neq 3}(m_r+n_r+k_r)}\;E_3\biggr)\nonumber\\
&&\qquad\qquad -\;\ln \biggl( \sqrt{\omega}^{\;- k_4 \cdots
    k_M\,z\,+\,\prod_{r\neq 3} (\cdots)}\; E_3\biggr)\biggr)\;\cdot\;
\sqrt{\omega}^{\;(m_3+n_3)\cdot \prod_{r\neq 3} (\cdots)}\nonumber\\
&&\qquad\qquad \cdot \; E_1^{m_1+n_1}\;E_2^{m_2+n_2}\; E_3^{m_3+n_3}\;
E_4^{m_4+n_4+k_4}\; \cdots \; E_M^{m_M+n_M+k_M}\nonumber \\
&& {z\;:=\;(\vec m, \vec n, \vec e_3,\cdots, \vec e_M)\;=\; m_1\,n_2 -
  m_2\,n_1 \choose \vec k\;=\; (0,\;0,\;k_3,\cdots,
  k_M)\qquad\qquad\qquad\qquad}\nonumber\\ 
&=& (\ln \omega)\;p_{i\vec m}\;x_{i\vec n}\;z\;(\vec m, \vec
n)\;\sqrt{\omega}^{\;\prod_1^M (m_r+n_r)}\; E_1^{m_1+n_1} \cdots
E_M^{m_M+n_M}\nonumber\\
&=& (m_1\;n_2 - m_2\;n_1)\;p_{i\vec m}\;x_{i\vec n}\;(\ln
\omega)\;\cdot\;T_{\vec m+\vec n}\nonumber\\
&&{\rm where}\ ({\rm for}\ r>3) - \sum_{n=1}^\infty
\sum_{k=0}^n {n\choose k}\;\frac{(-)^k}{n}\;
k\;\cdot\;E_r^k\;\cdot\;(\omega^{\cdots})^k\;=\;E\ {\rm was\ used.}\nonumber
\end{eqnarray}

However, 
\begin{equation}
c_{\vec m\vec n}\;=\;(\vec m, \vec n,\ {\rm anything})
\end{equation}
does \ub{not} satisfy (41). Moreover, even if one considers more
general deformations of the Hamiltonian, i.e. replacing the
sine-function in (41) by an arbitrary odd (power-series) function $f$
of the determinant, the corresponding condition, 
\begin{eqnarray}
&& f\,(\vec a_1, \cdots, \vec a_M)\;f\,(\vec a_1 + \cdots + \vec a_M,\;
\vec a_2',\cdots, \vec a_M')\;\cdot\c (\vec e, \vec a_1',\cdots \;
)\;=\;0 \nonumber\\
&& \qquad\qquad +\; (M\;\cdot\;2^M - 1)\;{\rm permutations}\;,
\end{eqnarray}
$\vec e = \displaystyle\mathop{\Sigma}_{r=1}^M (\vec a_r + \vec
a_r')$, \ can never be satisfied by any non-linear function $f$ -- as
on can see, e.g., by choosing $\vec a_r' = \mu_r \vec a_r$. \
Supposing $f(x) = \alpha x + \beta x^{2n+1} = \cdots \ ,$ and denoting
$(\vec a_1, \cdots, \vec a_M)$ by $z$, $\displaystyle\mathop{\Pi}_{r=1}^M
\mu_r$ by $\mu$, the terms $\mu_1, \alpha\,z\,\beta\ (\mu\,z)^{2 n+1}$,
\ e.g., (occurring only twice, with the same sign) could never
cancel. 

The preceding arguments possibly suffice to prove that, independent of
the above dynamical context, the Lie algebra of volume-preserving
diffeomorphisms of $T^{M>2}$  does not
possess any non-trivial
deformations.\renewcommand{\thefootnote}{\fnsymbol{footnote}}\footnote[1]{{\normalsize
    M.
    Bordemann has informed me that apparently an even more general
    statement of this nature has recently been proven in
    [19].}}\renewcommand{\thefootnote}{\arabic{footnote}}

\vspace{1cm}

\section*{5. Rigidity of Canonical Nambu-Poisson Manifolds}

\noindent For the multilinear antisymmetric map (4), and $2M-1$
arbitrary functions $f_1, \cdots, f_{2M-1}$, one has (cp.~[5]):
\begin{eqnarray}
&&\{\{ f_M, f_1, \cdots, f_{M-1}\},\; f_{M+1}, \cdots ,
\;f_{2M-1}\}\nonumber\\
&&\quad +\; \{ f_M, \{ f_{M+1}, f_1, \cdots, f_{M-1}\}, f_{M+2},
\cdots, f_{2M-1}\}\nonumber\\
&&\quad +\;\cdots\; +\; \{ f_M, \cdots f_{2M-2}, \{ f_{2M-1}, f_1,
\cdots, f_{M-1}\}\}\nonumber\\
&&\quad =\;\{\{ f_M, \cdots, f_{2M-1}\}, f_1, \cdots, f_{M-1}\}\;.
\end{eqnarray}
Takhtajan [5], stressing its relevance for time-evolution in
Nambu-mechanics [1], named (50) `Fundamental Identity (FI)', and
defined a `Nambu-Poisson-manifold of order $M$ ' to be a manifold $X$
together with a multilinear antisymmetric map \ $\{ \cdots \}$ \
satisfying (50) and the Leibniz-rule 
\begin{equation}
\{ f_1\tilde f_1,f_2,\cdots, f_M\}\;=\;f_1\{ \tilde f_1, f_2,\cdots,
f_M\}\;+\;\{f_1, \cdots, f_M\}\;\tilde f_1
\end{equation}
for functions $f_r : X \to \mathbb{R}$ (or $\mathbb{C}$).

Without (51), i.e. just demanding (50) for an antisymmetric $M$ linear
map: $V\times \cdots\times V\to V$, $V$ some vectorspace, Takhtajan
defines a `Nambu-Lie-gebra' [5], -- also called `Fillipov [6] Lie
algebra' [7]). I would like to point out various other
identities satisfied by canonical Nambu-Poisson brackets (4), and show
that all of them -- including (50)! -- do \ub{not} allow deformations
(of certain natural type), if \ $M>2$. 

At least from a non-dynamical point of view, all identities involving
Nambu-brackets obtained
from antisymmetrizing the product of two determinants formed from $2M$
\ $M$-vectors, 
\begin{equation}
(\vec a_1 \cdots \vec a_M) (\vec a_{M+1} \cdots \vec a_{2M})
\end{equation}
with respect to $M+1$ of the $\vec a_\alpha (\alpha = 1 \cdots 2M)$ \
should be treated on an equal footing. \ For $M=3$, e.g., one has --
apart from
\begin{eqnarray}
&& (\vec a\; \vec b \;\vec c_1)(\vec c_2\; \vec c_3 \;\vec c_4)\;-\;(\vec a\;
\vec b\; \vec c_2) (\vec c_3\; \vec c_4 \;\vec c_1) \nonumber\\
&&+\; (\vec a \;\vec b\; \vec c_3) (\vec c_4\; \vec c_1 \;\vec c_2)\;-\; (\vec
a \;\vec b \;\vec c_4) (\vec c_1\; \vec c_2 \;\vec c_3)\;=\; 0\;,
\end{eqnarray}
which gives rise to (50)$_{M=3}$ for functions \ $f \,\in\,T^3$
 -- also 
\begin{equation}
(a\;\vec c_{[1} \;\vec c_2)(\vec c_3\;\vec c_{4]}\; \vec b)\;=\; 0\;,
\end{equation}
yielding the following 6-term identity (FI)$_6$ \ (which can of course also be
proven by using just the definition (4), \ $\{ f, g, h\} =
\epsilon_{\alpha\beta\gamma}\ \partial_\alpha \; f\; \partial_\beta \;
g\; \partial_\gamma\;h$, \ rather than (54); i.e.~not necessarily
specifying the manifold $X$):
\begin{equation}
\{\{ f,\; f_{[1},\; f_2\}\;f_3,\; f_{4]}\}\;=\;0
\end{equation}
as well as the 4-term identity $(\widetilde{{\rm FI}})$, 
\begin{eqnarray}
&&\quad \{\{ f, f_1, f_2\},\;g,f_3\}\nonumber\\
&&+\; \{\{ f, f_2, f_3\},\;g,f_1\}\\
&&+\;\{\{ f, f_3, f_1\},\;g,f_2\}\;=\;-\;\{f, g, \;\{ f_1, f_2,
f_3\}\}\nonumber 
\end{eqnarray}
- -- each of which is independent of (50)$_{M=3}$ \ (while any 2 of the
3 identities yield the 3$^{{\rm rd}}$).

Naively, one would think that (56) should follow from (50)$_3$ alone,
as (54) follows from (53) \ (perhaps one should note that for general
$M$, a theorem concerning vector invariants [8] states,
that any (!) vector-bracket identity is an algebraic consequence of 
\[
( \vec a_{[1\;} \vec a_2 \;\cdots\; \vec a_M)\;(\vec
a_{M+1]\;}\cdots\;\vec a_{2M})\;=\;0\; ;
\]
however, in the proof of (56) via vector-bracket identities, one in
particular needs (54) for the special case $\vec a = \vec b$ -- which
cannot be stated as an identity between functions on $X$ .)
Curiously (with respect to a statistical approach to $M$-branes),
vector-bracket identities (`Basis Exchange Properties' [9]) also play
an important role in combinatorical geometry. 


>From an aesthetic point of view, the most natural quadratic identity
for (4) is  
\begin{equation} 
\sum_{\sigma\,\in\,S_{2M-1}} \;({\rm
  sign}\;\sigma)\{\{f_{\sigma 1},\cdots , f_{\sigma M}\}\; f_{\sigma
  M+1},\cdots, f_{\sigma 2M-1}\}\;=\;0\; .
\end{equation}
For $M=3$, e.g., one could see this to be a consequence of (50)$_3$
and (56) by grouping the 10  distinct terms in (57)
according to whether \ $\{ f_{\sigma 1}, f_{\sigma 2}, f_{\sigma 3}\}$
\ contains both $f_4$ and $f_5$ (3 terms, `type A'), 
only one of them (3 `B-terms' and 3 `C-terms') or none of them (1
term, `type D'); for the B (resp.~C)-terms one can use (56) while (50)
for the A-terms, to get \ 
$\pm\;\{ f_4, f_5, \{ f_1\,f_2\,f_3\}\}$ \ for each of the 4 types,
and for the B and C-terms with a sign opposite to the one obtained
from the D (and A) term(s). \ (57) (taken without the
derivation-requirement) is a beautiful generalisation of Lie-algebras
$(M=2)$, and has recently started to attract the attention of
mathematicians -- mostly under the name of $(M-1)$-ary Lie algebras
[10 - 13].
\renewcommand{\thefootnote}{\fnsymbol{footnote}}\footnote[1]{
{\normalsize I would like to thank W. Soergel for mentioning
  refs. [10]/[11] to me and J.L. Loday for sending me a copy of [10]
  and [12]; also, I would like to express my gratitude to
  R.~Chatterjee and L. Takhtajan for sending me their papers on Nambu
  Mechanics (cp.~[5]).}}\renewcommand{\thefootnote}{\arabic{footnote}} 

Unfortunately, all identities (50), (55)--(57), can be shown to be
rigid, in the following sense: assuming that
\begin{equation}
[ T_{\vec m_1},\cdots, T_{\vec m_M} ]_\lambda\;=\; g_\lambda\;(( \vec m_1,
\cdots, \vec m_M))\;T_{\vec m_1 +\cdots + \vec m_M}
\end{equation}
with $g_\lambda (x)$ a smooth odd function proportional to \ $x +
\lambda^n\,c\,x^n$ \ as \ $\lambda \to 0$ $(n>1)$ \ any of the above
identities will require the constant $c$ to be equal to zero (I have
proved this only for $M=3$, and in the case of (57) -- the a priori
most promising case -- for general $M>2$). 

Concerning
\begin{eqnarray}
&&\; g_\lambda \left( (\vec a, \vec b, \vec c_1)\right)\; g_\lambda
\left( (\vec a + \vec b + \vec c_1, \vec c_2, \vec c_3
  )\right)\nonumber \\
&+&\; g_\lambda \left( (\vec a, \vec b, \vec c_2)\right)\;g_\lambda
\left( ( \vec a + \vec b + \vec c_2, \vec c_3, \vec c_1)\right)
\nonumber\\
&+&\; g_\lambda \left( (\vec a, \vec b, \vec c_3)\right)\;g_\lambda
\left( ( \vec a+\vec b +\vec c_3, \vec c_1, \vec c_2)\right)
\nonumber\\
&\stackrel{\textstyle!}{=}&\; g_\lambda \left( (\vec c_1, \vec c_2,
\vec c_3)\right)\; g_\lambda \left( (\vec c_1+\vec c_2 +\vec c_3,
\vec a,\, \vec b)\right)\; ,
\end{eqnarray}
i.e. the deformation of (50)$_{M=3}$, one could assume \ $z := (\vec
c_1, \vec c_2, \vec c_3) \neq 0$, $\vec a =
\sum_1^3 \ \alpha_r \vec c_r$, $\vec b =
\sum_1^3 \beta_r \vec c_r$, such that $g
  (y) := \bar g_\lambda (y) := g_\lambda (zy)$ must satisfy 
\begin{eqnarray}
&& g\; (\alpha_2 \,\beta_3 - \alpha_3\,\beta_2)\;\cdot\;g \;(1 + \alpha_1 +
\beta_1)\nonumber \\
&& \quad +\  {\rm cyclic \ permutations} \\
&& \quad = \;g\; (1)\;\cdot\;g\, (\alpha_2\,\beta_3 -
\alpha_3\,\beta_2\,+\,{\rm cycl.})\nonumber
\end{eqnarray}
for all $\alpha_r, \beta_r$; which is clearly impossible for any
nonlinear $g$ of the required form, (e.g., as in next to lowest order
in $\lambda$ the terms $\alpha_1 (\alpha_2\,\beta_3)^{n>1}$ appear
only once).

Similarly, the deformation of (56) is impossible due to the analogous
requirement 
\begin{eqnarray}
&& g\,(\alpha_3)\;g\,\left(
  \beta_2-\beta_1\,+\,(\alpha_1\,\beta_2\,-\,\alpha_2\,\beta_1)\right)
\;+\;{\rm cycl.}\nonumber\\
&&\qquad\quad \stackrel{\textstyle!}{=}\;-\, g\,(1)\;g\,\left( (
  \alpha_1\,\beta_2\,-\, \alpha_2\,\beta_1)\;+\; {\rm cycl.}\right)\; .
\end{eqnarray}
Finally, concerning possible deformations of (57), let $( \vec a_1,
\cdots, \vec a_M) \neq 0$, and 
\begin{eqnarray}
&&\vec a_{M+\bar r}\;=\;\sum_{s=1}^M\;\alpha_s^{(\bar r)}\;\vec a_s \
(\bar r = 1, \cdots, M-1);\nonumber\\
&&{\rm then \ } g\; ( 1+\alpha_1^{(1)} +\cdots +
\alpha_1^{(M-1)})\;\cdot\; g\;
\left( \underbrace{\begin{array}{ll}
1 & \\
0 & \vec\alpha^{\;(1)} \cdots \vec\alpha^{\;(M-1)}\\
\vdots & \\
0 & 
\end{array}}\right)\; , \nonumber\\
&&\phantom{ then \ g\; ( 1+\alpha_1^{(1)} +\cdots +
\alpha_1^{(M-1)})\;\cdot\; g\;Zeichnung }=:\;[1] \nonumber
\end{eqnarray}
e.g., contains (in next to lowest order in $\lambda$) a term \
$\alpha_1^{(1)}\,\cdot\,\alpha_1^{(2)}\,\cdot\,[1]$ \ (of total degree
$(M+1)$ in the $\alpha_s^{(\bar r)}$), which cannot appear anywhere
else (in the same order in $\lambda$), -- in contradiction to the
assumption that (57) should hold for $[\cdots]_\lambda$ \ (cp.~(58))
replacing the curly bracket (4). 

\vspace{1cm}

\vfill\eject

\section*{6. A Remark on Generalized Schild Actions}

\noindent Consider
\begin{equation}
S \;:=\;-\;\int d\varphi^0\;d^M\varphi\;f(G)\ , 
\end{equation}
where \ $G := (-)^M$ det$\;(G_{\alpha\beta})$, $G_{\alpha\beta} :=\
\frac{\partial x^\mu}{\partial\varphi^\alpha} \ \frac{\partial
  x^\nu}{\partial\varphi^\beta}$ $\eta_{\mu\nu}$, \ $\eta_{\mu\nu}$ =
diag $(1,-1,\cdots,-1)$,\break
 $\alpha,\beta=0,\cdots, M$ \ and $f$ some
smooth monotonic function like $G^\gamma$ $(\gamma=1$ resp. $\frac 1 2$
\ corresponding to a generalized Schild-, resp. Nambu-Goto, action for
$M$-branes). Apart from a few subtleties (like $\gamma=1$ allowing for
vanishing $G$, while $\gamma = \frac 1 2$ does not) actions with
different $f$ are equivalent, in the sense that the equations of
motion,
\begin{equation}
\partial_\alpha\,\left( f' (G) G\,G^{\alpha\,\beta}\;\partial_\beta
  \;x^\mu\right)\;=\; 0 \qquad \mu\;=\;0\cdots\cdot D-1
\end{equation}
are easily seen to imply
\begin{equation}
\partial_\alpha\;G\;=\;0 \qquad \alpha\;=\;0,\cdots,M
\end{equation}
(just multiply (63) by $\partial_\epsilon x_\mu$ and sum) -- unless
$f(G)$ = const. $\sqrt{G}$, in which case (62) is fully
reparametrisation invariant and a parametrisation may be assumed in
which $G$ = const. (such that (63) becomes proportional to
$\partial_\alpha (G^{\alpha\beta} \ \partial_\beta x^\mu)$ also in
this case). Due to
\begin{equation}
G\;=\;\sum_{\mu_1<\cdots <\mu_{M+1}}\;\{ x^{\mu_1}, \cdots,
x^{\mu_{M+1}}\}\; \{ x_{\mu_1},\cdots, x_{\mu_{M+1}}\}
\end{equation}
(63) may be written as (cp. [14] for strings, and [15] for membranes,
in the case of $\gamma = 1$ resp. $\frac 1 2$)
\begin{equation}
\left\{ f'(G) \{ x^{\mu_1},\cdots,x^{\mu_{M+1}}\},\ x_{\mu_2},\cdots,
  x_{\mu_{M+1}}\right\} \;=\;0 \ ,
\end{equation}
whose deformed analogue (note the similarity between $G$ = const. and
condition (3.9) of [16])
\begin{equation}
\left[ [ x^{\mu_1}, \cdots, x^{\mu_{M+1}}], \ x_{\mu_2},\cdots,
  x_{\mu_{M+1}}\right]\;=\;0 
\end{equation}
looks very suggestive when thinking about space-time quantization in
$M$-brane theories.

\vspace{1cm}

\section*{7. Multidimensional Integrable Systems from\\
$\phantom{7.\;}$ M-algebras}

\noindent Several ideas used in the context of integrable systems are
based on bilinear operations. Our problems to extend results about low
(especially 1+1) dimensional integrable field theories to higher
dimensions may well rest on precisely this fact. Already some time
ago, attempts were made to overcome this difficulty by generalizing
Lax-pairs to -triples ([3],~p.~72) and Hirota's bilinear equations for
`$\tau$-functions' [17] to multilinear equations ([3],~p.~107-111). 

At that time, good examples were lacking, and -- not being an
exception to the rule that generalisations involving the number of
dimensions (of one sort or an other) are usually hindered by
implicitely low dimensional point(s) of view -- the proposed
generalisation of Hirota-operators may have still been too naive;
while hoping to come back to the question of multidimensional
$\tau$-functions in the near future, I would now like to give an
example ($M>3$ will then be obvious) for an equation of the form
\begin{equation}
\dot{\cal{L}}\;=\;\frac 1\rho\;\{ {\cal{L}},\;{\cal{M}}\,_1,\; {\cal{M}}\,_2 \}
\end{equation}
being equivalent to the equations of motion of a compact 3
dimensional manifold $\widehat{\sum} \subset \mathbb{R}^4$ \ (described
by a time-dependent 4-vector \ $x^i
(\varphi^1,\varphi^2,\varphi^3,t))$, moving in such a way that its
normal velocity is always equal to the induced volume density \ $\sqrt{g}$ \
(on $\widehat{\sum}$) devided by 
a fixed non-dynamical density $\rho (\varphi)$ \ (`the' volume
density of the parameter manifold):
\begin{eqnarray}
\dot{x}_1 &=& \quad \frac 1\rho \; \{ x_2, x_3, x_4 \} \nonumber\\
\dot{x}_2 &=& - \frac 1\rho \; \{ x_3, x_4, x_1 \} \nonumber\\
\dot{x}_3 &=& \quad \frac 1\rho \; \{ x_4, x_1, x_2 \} \nonumber\\
\dot{x}_4 &=& - \frac 1\rho \; \{ x_1, x_2, x_3 \} \ .  
\end{eqnarray}
With the curly bracket defined as before (cp.~(4)), it will be an immediate
consequence of (68) that
\begin{equation}
Q_n\; := \; \int_\sum \; d^3\varphi\; \rho(\varphi)\; {\cal{L}}^n
\end{equation}
is time-independent (for any $n$). 

In [2] evolution-equations of the form (69) (in any number of dimensions) were shown
to correspond to the diffeomorphism invariant part of an integrable
Hamiltonian field theory (as well as to a gradient flow); one way to
solve such equations is to note ([18], [2]) that the time at which the
hypersurface will pass a point $\vec x$ in space will simply be a
harmonic function. 

In any case,  the (a) form of (${\cal{L}}, {\cal{M}}_1,
  {\cal{M}}_2$) that will yield the equivalence of (69) with (68) is:
\begin{eqnarray}
{\cal{L}} &=& (x_1 + ix_2) \frac 1\lambda\;+\; (x_3+ix_4) \frac
1\mu\;+\; \mu (x_3-ix_4)\;-\; \lambda (x_1-ix_2)\nonumber \\
{\cal{M}}_1 &=& \frac \mu 2 (x_3-ix_4)\;-\;\frac{1}{2\mu} (x_3+ix_4)
\\
{\cal{M}}_2 &=& \frac{\lambda}{2}
(x_1-ix_2)\;+\;\frac{1}{2\lambda}(x_1+ix_2)\nonumber 
\end{eqnarray}  
(involving two spectral parameters, $\lambda$ and $\mu$). Surely, this
observation will have much more elegant formulations, and conclusions.

\vspace{1cm}

\section*{Acknowledgement}

\noindent I would like to thank M. Bordemann, A. Chamseddine,
J. Fr\"ohlich, D. Schenker and M.~Seifert for valuable discussions.

\vspace{.5cm}

\baselineskip=0.45cm

\section*{References}

\begin{description}
\item[[\ 1]] Y. Nambu; Phys. Rev. {\it D7} $\#$ 8 (1973) 2405.
\item[[\ 2]] M. Bordemann, J. Hoppe; `Diffeomorphism Invariant
  Integrable Field Theories and Hypersurface Motions in Riemannian
  Manifolds'; ETH-TH/95-31, FR-THEP-95-26.
\item[[\ 3]] J. Hoppe; `Lectures on Integrable Systems'; Springer-Verlag
  1992.
\item[[\ 4]] J. Hoppe; `Quantum Theory of a Massless Relativistic
  Surface'; MIT Ph.~D. thesis 1982 and Elem. Part. Res. J. (Kyoto)
  {\it 80} (1989) 145.
\item[[\ 5]] L. Takhtajan; Comm. Math. Phys. {\it 160} (1994) 295.\\
             R. Chatterjee; `Dynamical Symmetries and Nambu Mechanics';
  Stony Brook pre\-print 1995.\\
             R. Chatterjee, L. Takhtajan; `Aspects of Classical and Quantum
  Nambu Mechanics' \  (1995; to appear in Lett. Math. Phys.).
\item[[\ 6]] V.T. Filippov; `$n$-ary Lie algebras'; Sibirskii
  Math. J. {\it 24} $\#$ 6 (1985) 126 (in russian). 
\item[[\ 7]] P. Lecomte, P. Michor, A. Vinogradov; `$n$-ary Lie and
  Associative Algebras'; preprint 1994.
\item[[\ 8]] H. Weyl; `The Classical Groups'; 2$^{{\rm nd}}$ edition,
  Princeton University Press.
\item[[\ 9]] N. White; `Theory of Matroids'; Cambridge University Press
  1987.
\item[[10]] J.L. Loday; `La renaissance des op\'erades'; in
  S\'eminaire Bourbaki, expos\'e 792, Novembre 1994.
\item[[11]] V. Ginzburg, M.M. Kapranov; Duke Math. J. {\it 76} (1994)
  203.
\item[[12]] Ph. Hanlon, M. Wachs; `On Lie $k$-Algebras; preprint 1993.
\item[[13]] A.V. Gnedbaye; C.R. Acad. Sci. Paris, {\it t.~321},
  S\'erie I, p. 147, 1995.
\item[[14]] A. Schild; Phys. Rev. {\it D16} (1977) 1722.
\item[[15]] A. Sugamoto; Nucl. Phys. {\it B215} [FS7] (1983) 381.
\item[[16]] S. Doplicher, K. Fredenhagen, J. Roberts;
  Comm. Math. Phys. {\it 172} (1995) 187.
\item[[17]] R. Hirota; `Direct methods of finding solutions of
  nonlinear evolution equations', in Lect. Notes in Math.~{\it 515},
  Springer-Verlag 1976.
\item[[18]] J. Hoppe; Phys. Lett. {\it B335} (1994) 41.
\item[[19]] P. Lecomte, C. Roger; `Rigidit\'e de l'alg\`ebre de Lie
  des champs de vecteurs unimodulaires'; Universit\'e IGD Lyon 1,
  preprint (1995).
\end{description}

\end{document}